\documentclass[
 aip, amsmath,amssymb,
reprint
]{revtex4-2}

\usepackage{graphicx}
\usepackage{mathptmx}

\newcommand{\tep}{\tau_{\rm ep}}
\newcommand{\cool}{\ell_{\rm E}}

\begin{document}
\title{The heating and cooling of 2D electrons at low temperatures} 

\author{A.~K. Jain}
\author{J.~T. Nicholls}\email{james.nicholls@rhul.ac.uk}
\affiliation{Physics Department, Royal Holloway, University of London, Egham
TW20 0EX, United Kingdom}
\author{S. N. Holmes}
\affiliation{Department of Electronic and Electrical Engineering, University College London, Torrington Place, London WC1E 7JE, United Kingdom}
\author{G. Jaliel}
\author{C. Chen}
\affiliation{Cavendish Laboratory, University of Cambridge, JJ Thomson Avenue, Cambridge, CB3 0HE, United Kingdom}
\author{I. Farrer}
\affiliation{Cavendish Laboratory, University of Cambridge, JJ Thomson Avenue, Cambridge, CB3 0HE, United Kingdom}
\address{Department of Electronic and Electrical Engineering, 
University of Sheffield, Mappin Street, Sheffield S1 3JD, United Kingdom}
\author{D.~A. Ritchie}
\affiliation{Cavendish Laboratory, University of Cambridge, JJ Thomson Avenue, Cambridge, CB3 0HE, United Kingdom}

\date{\today}
 
\begin{abstract}
We present measurements of the cooling length $\cool$ for hot electrons in a GaAs-based high mobility two-dimensional electron gas (2DEG). 
The thermal measurements are performed on a long 60~$\mu$m-wide channel, 
which is Joule-heated at one end, along which there are three similar hot-electron thermocouples, spaced $30~\mu$m apart. 
The thermocouples measure an exponentially decaying temperature profile
with a characteristic length $\cool$, which decreases from 23 to 16~$\mu$m 
as the lattice temperature increases from 1.8 to 5~K.
From a simple one-dimensional model of heat diffusion,  
we measure an inelastic scattering time which decreases from $\tau_i \approx 0.36$ to 0.18~ns.    
The measured $\tau_i$ has a magnitude and temperature dependence consistent with    
acoustic phonon scattering times.  
We discuss how the sample design can be varied for further thermal investigations. 
Knowledge of the temperature profile and its gradient
will prove useful in measurements of the thermal conductivity and the Nernst effect. 
\end{abstract}

\maketitle

The inelastic lifetime $\tau_i$ of an electron is an important quantity, 
as it determines how electrons in a device can dissipate their excess energy.
Also, in quantum devices, the inelastic length scale is believed
to provide the upper limit for the phase-breaking length.\cite{Lin2002}
In a GaAs-based two-dimensional electron gas (2DEG) of high electron mobility ($> 2 \times 10^6$~cm$^2$/Vs),
the resistivity of a 2DEG increases linearly with temperature for $T \gtrsim 2~$K;  
a property that is determined by the thermal population of acoustic phonons available for scattering with the electrons.  
In thermal studies of 2DEGs, heated electrons lose their excess energy to the lattice by two routes: by inelastic electron-phonon 
scattering following a $T^5$ dependence in the Bloch-Gruneisen regime, or by a $T^2$  cooling by the Ohmic contacts.  
A crossover between the two was observed \cite{appl98a} using 1D electron thermometers at 0.3~K, 
and at lower temperatures using current noise thermometry. \cite{Levitin2022}
In this work, we measure the thermal diffusion length using non-local heating; 
this directly gives the inelastic time $\tau_i$,  which we find to be determined by electron-phonon scattering.  

In this work, the electrons are Joule heated by passing a current through the 2DEG.  
As the electron-electron relaxation length is much less than the sample dimensions, the electrons equilibrate to a local electron 
temperature $T_e$ above the lattice temperature $T_L$, with a distribution that obeys Fermi-Dirac statistics.  
Previous electron thermometers in 2DEGs have been based on the temperature dependence 
of quantum corrections to the conductivity,\cite{Syme1989}
Johnson noise\cite{Kurdak1995}, and the thermoelectric properties of 2D bar gates,\cite{Chickering2009} 	
1D\cite{appl98a} and 0D devices.\cite{Mavalankar2013}            
Chickering \mbox{\it et al.}\cite{Chickering2009} showed that an
accurate hot-electron thermocouple (HET) can be created from a pair of bar gates 
which can measure the temperature $T_e$ of the electrons in the channel 
when they are heated by as little as 10~mK above the lattice temperature $T_L$. 
At low temperatures, the thermovoltage $V_{th}$ developed across a single bar gate of the HET is given by 
\begin{equation}
 V_{th} =- S^{d} \,  (T_e-T_L),  
\end{equation}  
where $S^{d}$ is the diffusion thermopower, which 
can be related to the electrical conductance $G$ of a single bar gate 
using the Mott-Cutler relation\cite{MC1969}
\begin{equation}
 S^{d} = -\frac{\pi^2 k_B^2 }{3 e}  T  
 \left (  \frac{\partial \ln G}{\partial E} \right )_{E=E_F},
 \label{MC}
 \end{equation}
where the energy derivative is determined at the Fermi energy $E_F$ of the 2DEG. 
In the supplementary information (SI) we show a typical calibration using Eq.~\ref{MC} of a single bar-gate; 
$\frac{\partial \ln G}{\partial E_F}$ is determined from $\frac{d \ln G}{d V_g} \frac{dV_g}{dE_F}$, 
where $G(V_g)$ is the measured conductance characteristic of the bar gate as a function of gate voltage $V_g$.  

The diffusion thermopower $S^d$ of a single bar gate can be written\cite{Chickering2009} as  
\begin{equation}
S^{d} = -\frac{\pi^2 k_B}{3 e} \frac{T}{T_F} (1 + \alpha),
\label{eq:Sd}
\end{equation} 
where $T_F$ is the Fermi temperature (proportional to $n$). 
The quantity $\alpha$ is the elastic scattering parameter given by \mbox{$\alpha = d( \ln \tau_e) / d (\ln n)$}, 
describes the energy-dependent properties of the 2DEG, 
where $\tau_e$ is the elastic scattering time and $n$ is the 2D electron density.

Devices were fabricated from two similar wafers (C2681 and T612) grown by molecular beam epitaxy.
In both wafers the 2DEG is created 250~nm below the sample surface at the
Al$_{0.33}$Ga$_{0.67}$As/GaAs interface; on top of the undoped GaAs
there is 240~nm of Al$_{0.33}$Ga$_{0.67}$As, capped with a 10~nm GaAs top layer. 
There is Si-doping in the upper 200~nm of AlGaAs, giving a spacer layer distance of 
40~nm between the dopants and the 2DEG.  
The main results presented in the figures is Device~1 from C2681
in which the mobility and carrier density at 4.2~K are  
$\mu =2.8 \times 10^{6}$~cm$^{2}$/Vs and  
$n= 1.5 \times 10^{11}$~cm$^{-2}$, 
with a resistance per square of $R_{sq} \approx 15~\Omega$ and a diffusivity $D= 1.44$~m$^2$/s.
The elastic mean free path is $18~\mu$m, and from the 
variation of $n$ and $\mu$ with gate voltage we determine $\alpha \approx 1$. 
The properties of all samples investigated will be presented in Table~\ref{Tab1}, later in the paper.  

\begin{figure}
\includegraphics[width=8.6cm]{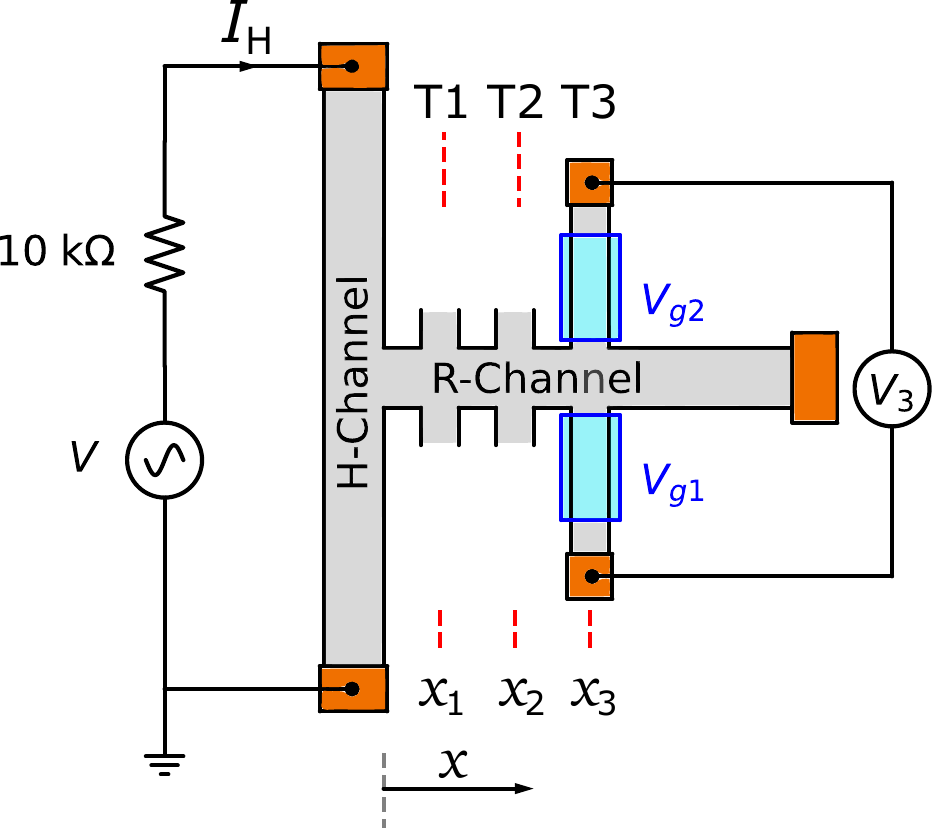}  
\caption{\label{fig1} 
Schematic of the T-shaped device, which consists of $60~\mu$m wide heating (H) 
and relaxation (R) channels, with lengths $720~\mu$m and 600~$\mu$m, respectively. 
All Ohmic contacts in the device (shown in orange)  
are thermally grounded to the lattice temperature $T_L$. 
In {\em non-local} measurements, an AC heating current of amplitude $I_H$ and frequency $f$ is driven between the Ohmic 
contacts on the H-channel; all other contacts in the device are electrically floating.
Due to heating by $I_H$, the electrons at the midpoint of the H-channel at $x=0$ 
are heated above $T_L$ by an excess temperature $\Delta T_0$.
The excess temperature, $\Delta T(x)=T_e(x) - T_L$, along the R-channel decreases with increasing $x$,
and is measured using three similar hot-electron thermocouples (HETs)  
centered at $x_1=22.5~\mu$m, $x_2=52.5~\mu$m and $x_3=82.5~\mu$m.
Thermocouple T3  is shown in detail; it consists of two 15~$\mu$m wide voltage probes, 
in which the electron carrier densities $n_1$ and $n_2$ can be varied using 
voltages $V_{g1}$ and $V_{g2}$ applied to surface gates of length $L_g=550~\mu$m.  }  
\end{figure}

Figure~\ref{fig1} is a schematic of our T-shaped device which consists of two 60~$\mu$m wide channels, 
a heating (H) and a relaxation (R) channel.  
There are three similar HETs along the R-channel, 
positioned at $x_1, x_2$ and $x_3$,  with thermocouple T3 positioned at $x_3$ shown in detail. 
From Eq.~\ref{eq:Sd} the diffusion thermopower of T3 is $\Delta S^d \propto (1/n_{1} - 1/n_{2})$, 
where $n_{1}$ and $n_{2}$ are the 2D carrier densities under the 
two bar gates controlled by applied gate voltages $V_{g1}$ and $V_{g2}$.
Experiments show\cite{Chickering2009} that the measured thermopower $\Delta S$ of a HET 
follows the Mott-Cutler prediction (based on Eq.~\ref{MC}) for $\Delta S^d$ up to $T_L \approx 2$~K.  
At higher temperatures, $\Delta S= \Delta S^d+ \Delta S^g$, is enhanced by a phonon drag contribution $\Delta S^g$, 
such that $\Delta S^g \approx 0.35 \ \Delta S^d$  at $T_L=4$~K.  See Ref.~\onlinecite{Chickering2009} and the SI for further details. 

Using the Seebeck effect, we investigate the electron's excess temperature $\Delta T= T_e-T_L$ in both 
{\em local} and {\em non-local} measurements.  
When a heating current is passed along the R-channel the HETs act as local 
electron thermometers.\cite{Chickering2009}. 
The cooling length $\cool$ is determined from  {\em non-local} measurements of the 
excess electron temperature $\Delta T(x)$, using the set-up shown in Fig.~\ref{fig1}.
Electrons are heated in the H-channel 
and $\Delta T$ decays down the R-channel,  
as measured in previous studies.\cite{Syme1989, Bakker2012,Sierra2015, Narayan2016, Waissman2021}
Away from the heating current, the decay in $\Delta T(x)$ is characterized by a cooling length $\cool$.   
In thermal diffusion the electrons exhibit an energy relaxation length, or cooling length, 
given by $\cool= \sqrt{D \, \tau_i }$ where the $D$ is the diffusivity of the 2DEG. 
We will describe a method to measure $\cool$,  and hence $\tau_i$.        

In early work\cite{Syme1989} on T-shaped devices the temperature decay in a 
2D silicon inversion layer was used to determine its thermal conductivity.
Previous studies\cite{Billiald2015} in a high-mobility 2DEG 
had only one working HET positioned $150~\mu$m down the R-channel.
A maximum cooling length of $250~\mu$m was determined 
at 170~mK from the variations in the measured thermovoltage, 
as  the carrier density under one of the $150~\mu$m long bar gates was varied.
In contrast, we will determine $\cool$ at higher temperatures 
using measurements of $\Delta T(x)$ from three HETs that are spaced $30~\mu$m apart. 
Our bar gates are much longer ($L_g=550~\mu$m) than the measured $\cool$, thus simplifying the analysis. 

In the non-local heating set-up an AC current $I_H \sin (\omega t )$ of 
frequency $f = \omega/(2 \pi) = 64$~Hz  is passed through the H-channel. 
Due to Joule heating, electrons in the middle of the H-channel at $x=0$ 
will have an oscillating electron temperature  
\begin{equation}
T_e = T_L + \Delta T_0 \,  \left [1-\cos (2 \omega t ) \right ].
\label{e:temp}
\end{equation}
For small AC heating currents, when $T_L \gg \Delta T_0$, 
the electrons will have an excess temperature $\Delta T=T_e-T_L$, 
that oscillates with frequency $2f$ with a phase shift of $\pi/2$. 
It is assumed that the electrons in the large area ($250\mu$m$\times150\mu$m) 
\mbox{AuNiGe} Ohmic contacts, with a contact resistance of $\sim 50~\Omega$, 
are anchored to the lattice temperature $T_L$.
Due to heating of the electrons in the R-channel between the voltage probes of T3, 
a thermovoltage $V_3$ at frequency $2f$ is measured between the two Ohmic contacts. 

\begin{figure}
\includegraphics[width=8.5cm]{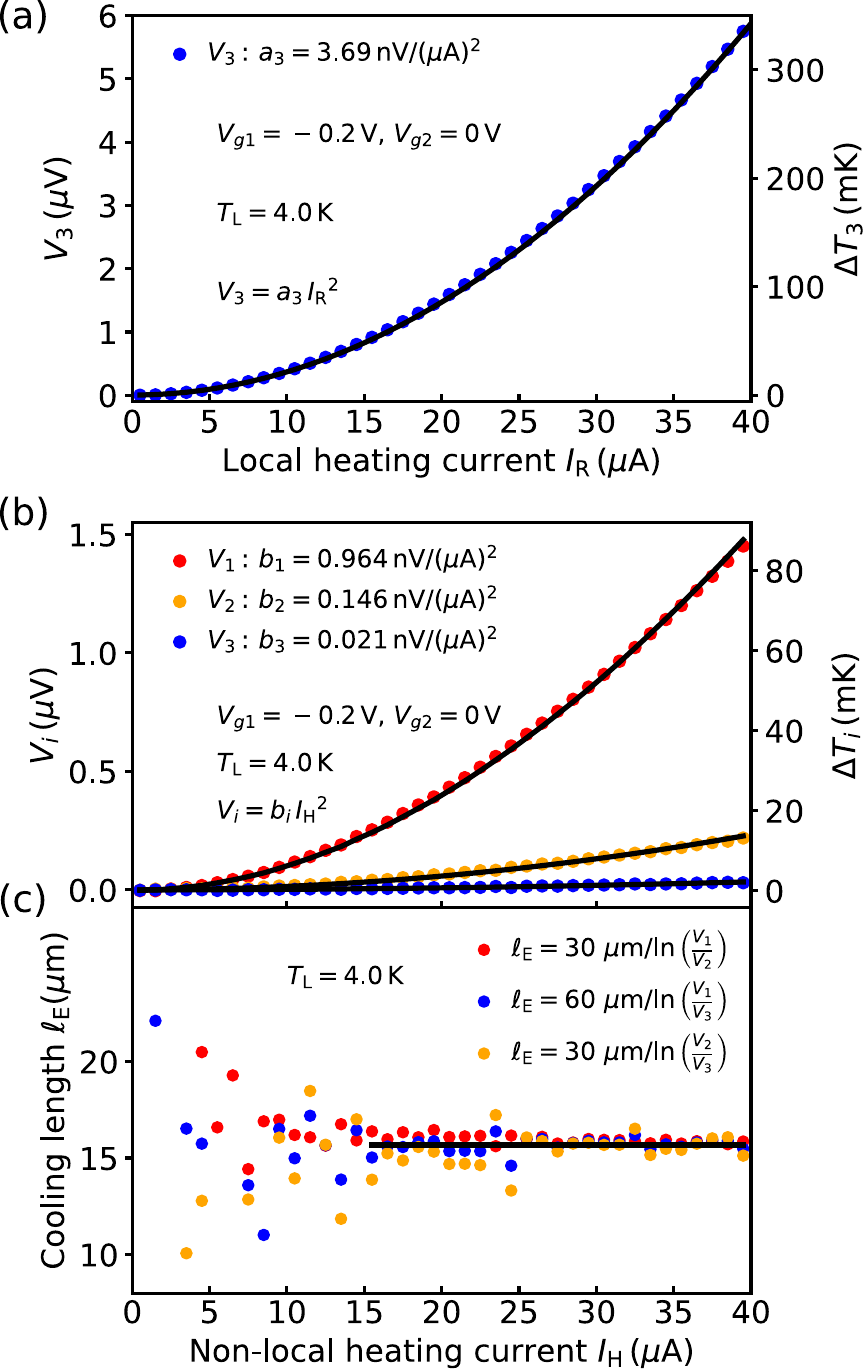}    
\caption{\label{fig2} Measurements of thermovoltages at $T_L=4.0$~K, 
when the gate voltages on all three HETs were set to $V_{g1}=-0.2$~V and $V_{g2}=0$~V. 
The Joule heating was either in the {\em local} or {\em non-local}
geometries shown in the insets; in both cases good fits were obtained to parabolic forms.  
(a)~In local measurements only the thermovoltage $V_3$ is plotted, 
with a parabolic fit $V_3 = a_3 I_R^2$ up to $40~\mu$A, where $a_3= 3.69$~nV/($\mu$A)$^2$.  
T1 and T2 have similar characteristics, with $a_1= 3.62$~nV/($\mu$A)$^2$ 
and $a_2= 3.47$~nV/($\mu$A)$^2$. 
(b)~In non-local measurements, the thermovoltages of the HETs 
show a decreasing response to $I_H$ as $x$ increases. 
(c)~The cooling length $\cool$ is determined from the data points in 
(b) using the three expressions in Eq.~\ref{len}. 
For $I_H > 20~\mu$A, the cooling length is $\cool = 15.8 \pm 0.3~\mu$m at 4.0~K. 
The estimated temperatures $\Delta T_i$ given on the right hand side of (a) and (b) are calculated 
using an electron-phonon cooling time of $\tep= 1 ~{\rm ns}/T ({\rm in \ K})$; see SI for details. }
\end{figure}

Figure~\ref{fig2}(a) shows the thermovoltage $V_3$ 
at  \mbox{$T_L=4.0$~K}, as a function of a {\em local} heating current $I_R$ passed through the R-channel. 
$V_3$ follows a quadratic dependence on heating current, $V_3= a_3\, I_R^2$,
where $a_3 = 3.69$~nV/($\mu$A)$^2$. 
Similar thermovoltage characteristics were obtained for the T1 and T2 thermocouples, 
which were operated at the same bar gate voltages as T3. 
The fitted values for $a_1, a_2$ and $a_3$ are within 6\% of each other, 
and have been used to fine-tune the calibrations of the HETs when they are 
used for relative thermometry measurements in the non-local geometry.  

In {\em non-local} measurements, the electrons in the H-channel are heated by a current $I_H$, 
which sets up a temperature gradient along the R-channel, 
driving diffusive heat flow in the $x$-direction. 
In a simple one-dimensional description,\cite{Rojek2014} the excess temperature 
$\Delta T(x)$ is governed by the differential equation:
\begin{equation} 
\kappa \ \frac{d^2 ( \Delta T(x))}{dx^2} - \frac{C_e \, \Delta T (x)}{\tau_i}=0,
\label{diff}
\end{equation}
where $\kappa$ is the thermal conductivity, and $C_e$ is the electron heat capacity.   
If the R-channel is much longer than $\cool$, the solution to Eq.~\ref{diff} is an exponential decay 
\begin{equation} 
\Delta T(x) = \Delta T_0  \, \exp (-x/\cool),
\label{sol}
\end{equation}
with prefactor $\Delta T_0$, the excess temperature at $x=0$, and 
characterized by the cooling length  \mbox{$\cool = \sqrt{ \kappa \, \tau_i/C_e}$}.   
This length can be written\cite{Rojek2014,Massicotte2021} as 
\begin{equation}
\cool = \sqrt{D \, \tau_i},
\label{ell}
\end{equation} 
where $D = v_F^2 \, \tau_e/2$ is the 2D diffusivity, 
$v_F$ is the Fermi velocity, and $\tau_e$ is the elastic scattering time. 

In non-local measurements the excess temperature $\Delta T(x)$ 
generates thermovoltages in the HETs along the R-channel. 
Figure~\ref{fig2}(b) shows non-local measurements of $V_1, V_2$ and $V_3$ at $T_L=4.0$~K;   
similar to local measurements there are good fits to the quadratic form $V_i= b_i \,  I_H^2$, 
with the values of $b_i$ given in the figure. 
The most important feature is the diminishing effect of $I_H$ on $V_i$ as $x$ increases. 
If a similar calibration for all the HETs is assumed from the local measurements in Fig.~\ref{fig2}(a), 
then the exponential decay in $\Delta T(x)$ described by Eq.~\ref{sol} 
will also be measured in the thermovoltages such that
\begin{equation}
V_i = \Delta S \, \Delta T (x_i) = \Delta S \, \Delta T_0 \, e^{-x_i/\cool},  
\label{prop}
\end{equation}
where $\Delta S$ is the magnitude of the thermopower of the three similar HETs.
From measurements of the ratio $V_j/V_i$, the cooling length $\cool$ can be obtained  
from  $\cool = (x_i-x_j)/\ln(V_j/V_i)$, which is derived from Eq.~\ref{prop}.
Three possible measurements of $\cool$ are: 
\begin{equation}
\cool = \frac{30~\mu {\rm m}}{\ln(V_1/V_2)},  \quad 
\cool = \frac{30~\mu {\rm m}}{\ln(V_2/V_3)},  \quad {\rm and}  
\quad \cool = \frac{60~\mu {\rm m}}{\ln(V_1/V_3)},
\label{len}
\end{equation}
which are plotted in Fig.~\ref{fig2}(c) as a function of $I_H$.  
For \mbox{$I_H <15~\mu$A} the $V_j/V_i$ ratios show fluctuations;  
at higher $I_H$ these diminish and the three plots settle to give $\cool= 15.8 \pm 0.3~\mu$m. 

\begin{figure}
\includegraphics[width=8.5cm]{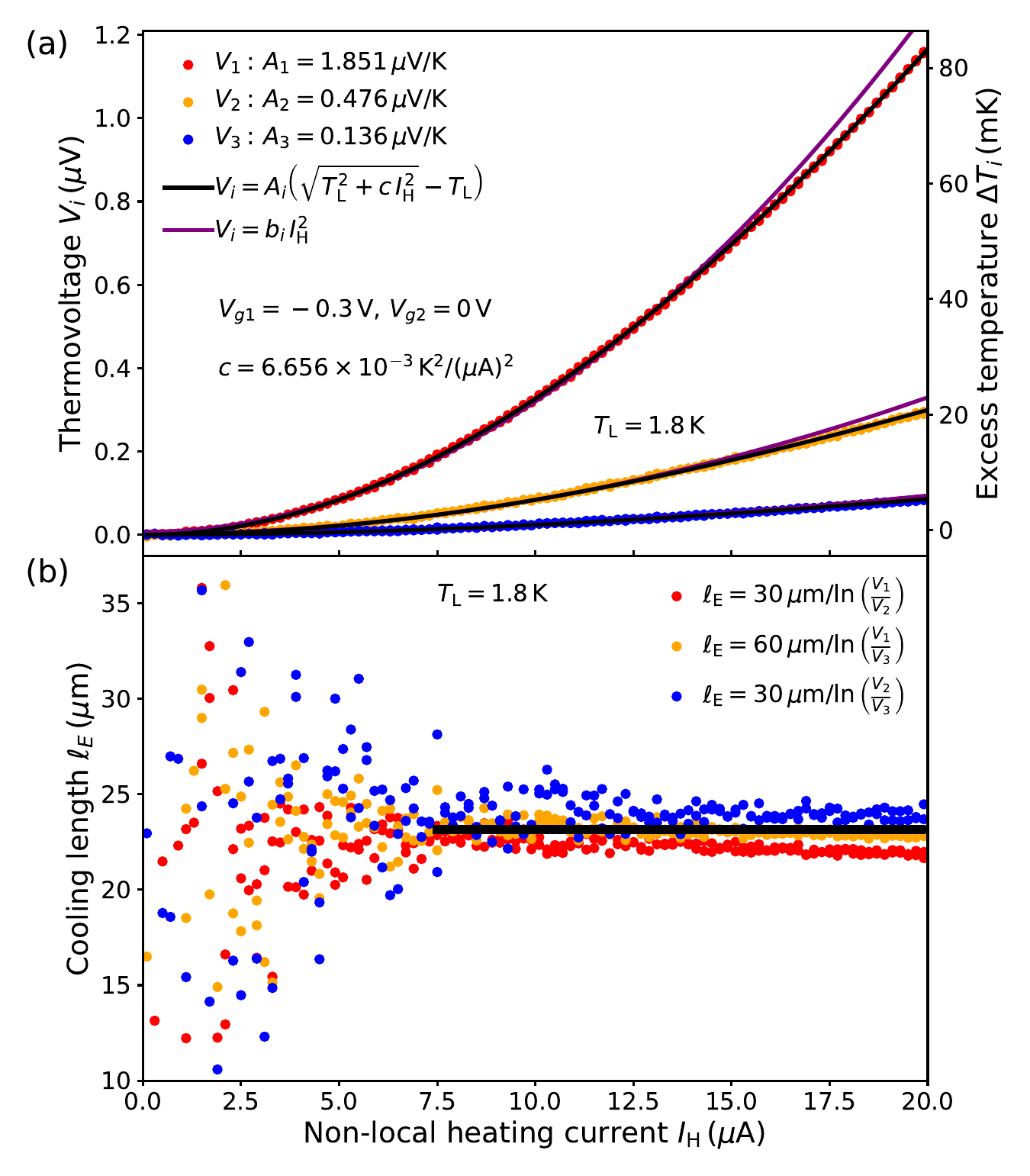}         
\caption{\label{fig3} (a) Non-local measurements of the thermovoltages $V_1$,
$V_2$, and $V_3$ at $T_L=1.8$~K as a function of the heating current $I_H$. 
Fits to Eq.~\ref{fitt} give the values of $A_i$ and $c$ shown in the figure.  
The excess temperatures $\Delta T_1$,  $\Delta T_2$ and $\Delta T_3$ 
on the right hand vertical axis, are calculated from 
calibrations based on the Mott-Cutler relation. 
(b)~The ratios of  $V_i/V_j$ as a function of  $I_H$ 
are used to determine a cooling length $\cool = 22.7 \pm 1.0~\mu$m at 1.8~K. }
\end{figure}

When the local and non-local measurements were 
performed at $T_L=1.8$~K, the electron-phonon 
cooling is diminished and the cooling length $\cool$ increases. 
At 1.8~K the HETs show similar characteristics in local heating 
and in Figs.~\ref{fig3}(a) and (b) we present  non-local measurements of $V_1, V_2$ and $V_3$, 
and the cooling length $\cool$ determined from their ratios. 
The behavior of $\Delta T(x)$ is not perfectly exponential, as shown by the   
$\sim 1~\mu$m difference in $\cool$ obtained from the $V_j/V_i$ ratios;  
the black horizontal line shows the cooling length $\cool = 22.7 \pm 1.0~\mu$m, which is 
determined from the average of the three ratios. 
As the temperature is further reduced (not shown) 
deviations from exponential behavior become more apparent, 
consistent with relatively stronger cooling by the Ohmic contacts 
as the electron-phonon interaction weakens.     

Due to the current $I_H$ flowing in the H-channel, one end of the R-channel is heated, and      
Figs.~\ref{fig2} and \ref{fig3} show an exponential decay of the 
excess temperature $\Delta T(x) \propto e^{-x/\cool}$,
characterized by the length scale $\cool$.  
The functional dependence of the three thermovoltages on $I_H$ are similar, being all determined 
by $\Delta T_0$, which is the excess temperature at the centre of the H-channel at the $x=0$.
At $T_L=4.0$~K, when electron-phonon cooling is strong, 
the electron temperature at the centre of a simple bar, well away from the Ohmic contacts,  
is constant, and $\Delta T_0$ is proportional to the Joule heating $I_H^2 R_{sq}$.  
At $T_L=1.8$~K the thermovoltages $V_i$ follow $I_H^2$ behavior for $I_H < 12~\mu$A, 
but for $I_H <20~\mu$A there is a better least-squares fit to
\begin{equation}
V_{i} =A_i \, \Delta T_0 = A_i   \left ( \sqrt{T_L^2 + c  \, I_H^2} - T_L \right ),
\label{fitt}
\end{equation}
a functional form used in hot-electron studies of diffusive metal wires.\cite{Henny1997,Huard2007}   
Figure~\ref{fig3}(a) shows the best fits, where $A_i$ and $c$ are fitting parameters. 
Using $c=0.0665$~(K/$\mu$A)$^2$ the fits of $V_1$, $V_2$ and $V_3$ to Eq.~\ref{fitt}  
are surprisingly good at $T_L=1.8$~K. 

\begin{figure}
\includegraphics[width=8.5cm]{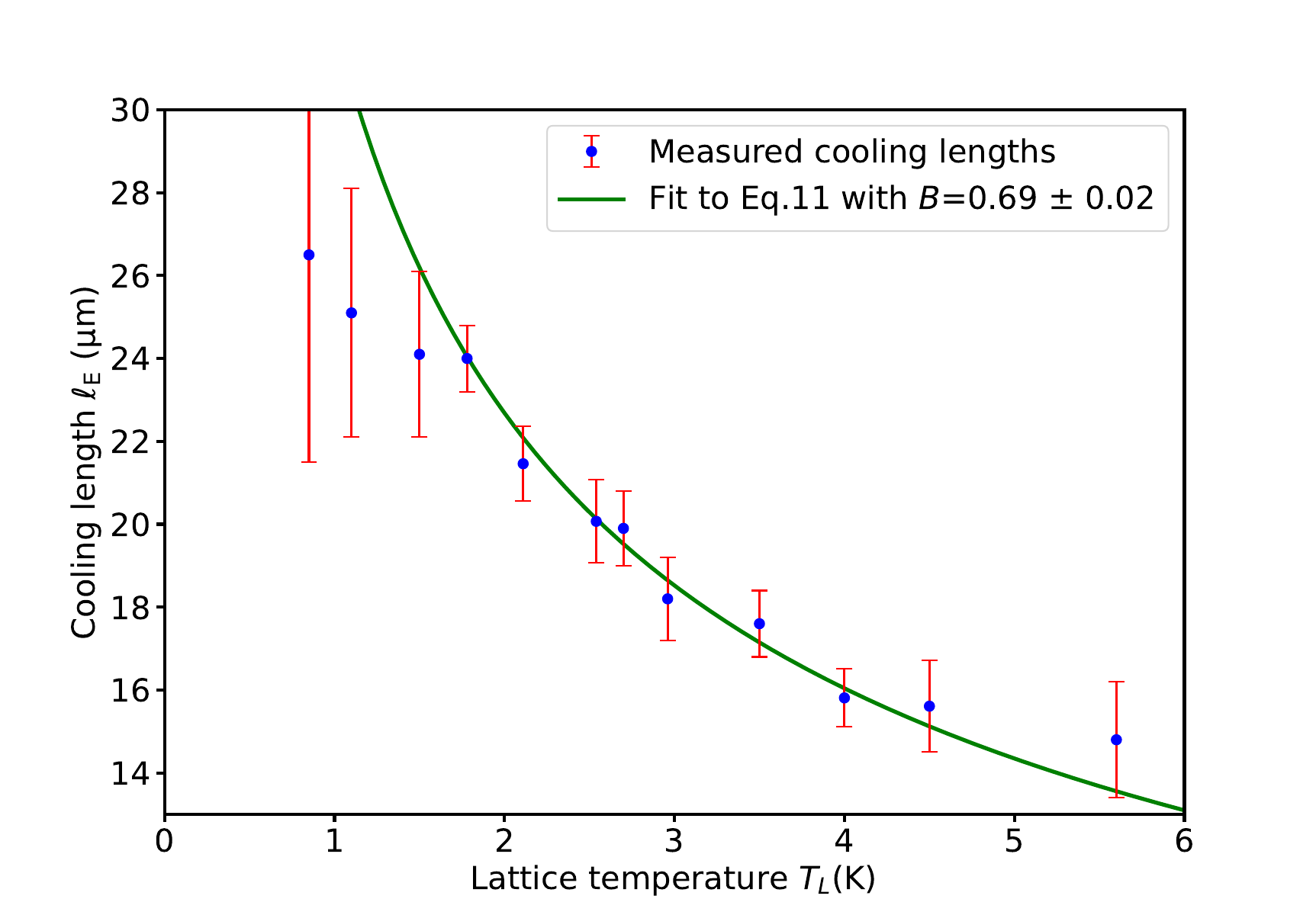} 
\caption{\label{fig4} 
The cooling length $\cool$ measured at different lattice temperatures $T_L$.    
Assuming that $\tau_i$ is equal to $\tep$ with the form given in Eq.~\ref{tau}, 
the best fit with $B= 0.69 \pm 0.02$ is obtained for $T_L \geq 1.8$~K.} 
\end{figure}

Figure~\ref{fig4} shows the length $\cool$ measured for 
different lattice temperatures $T_L$, when the gate voltages on all three HETs were set at 
$V_{g1}=-0.3$~V and $V_{g2}=0$~V.
At temperatures above 2~K, the electron-phonon mechanism is the expected source of inelastic scattering. 
It is well known that in GaAs-based 2DEGs 
the scattering from acoustic phonons gives rise to the linear temperature 
dependence of the resistivity $\rho(T)$ in high mobility 2DEGs for $T \gtrsim~$2~K.
To fit $\cool$ for $T_L> 1.8$~K, we have assumed that the inelastic time is equal to the electron-phonon scattering time:
\begin{equation}
\tau_i = \tep= \frac{B \times {\rm 1~ns}}{T  ({\rm in \ K})}, 
\label{tau}
\end{equation}
where the ${\rm 1~ns}/{\rm T  (in \ K)}$ term
comes from 
Ref.~\onlinecite{Arora1985}, assuming a 2DEG width of $d=10$~nm.
A best fit $B= 0.69 \pm 0.02$ is obtained for temperatures \mbox{$T_L \geq 1.8$~K}, 
giving electron-phonon scattering times similar to those obtained from $\rho(T)$ measurements.  

\begin{table*}[ht]
\begin{ruledtabular}
\begin{tabular}{|c|c|c|c|c|c|c|c|}
Device  & $n$ ($\times 10^{11}\, {\rm cm}^{-2})$ & $\mu$ ($\times 10^6\,\rm{cm^2/Vs}$)  & $W~(\mu$m) & 
$w~(\mu$m) &  $D$~(m$^2$/s)  & {$\cool \  (\mu\rm{m})$}  &   $\tau_i$ (ns)\\
\hline
1 & 1.50 & 2.82 &  60 & 15 & 1.51 & 15.7$\pm$0.3  &  0.163$\pm$0.006   \\    
\hline
2 & 1.50 & 2.91 & 60 & 15 & 1.58 & 16.1$\pm$0.2  &    0.164$\pm$0.004  \\ 
\hline
3 & 2.51 & 2.18 & 30 & 10  & 1.95 & 18.5$\pm$0.2 &   0.176$\pm$0.004   \\ 
\hline
4 & 2.66 & 2.27 &  60 &  5 & 2.16 & 20.0$\pm$0.5 &   0.185$\pm$0.009  \\ 
\hline
\end{tabular}
\caption{Properties of the four devices measured at $T_L=4.2$~K.  $w$ is the width of the 
voltage probes and $W$ is the width of the relaxation channel.  
Measurements from Device~1 are shown in Figs.~2-4. }
\label{Tab1}
\end{ruledtabular}
\end{table*}

Four different devices were investigated, with electrical and 
thermal results at $T_L=4.2$~K summarized in Table~\ref{Tab1}.   
Devices~1 and 2 were fabricated from wafer C2681 using the same mask set,  
where the heating channel and the relaxation channel are both $60~\mu$m wide.   
The measured $\tau_i$ at 4.2~K are very close for the two devices, 
and the value $B=0.687$ is consistent with measurements in Fig.~4.   
Devices 3 and 4 were fabricated from wafer T612, which has similar growth parameters to C2681,
using a mask set with a $60~\mu$m-wide heating channel, 
but with different widths for the voltage probes ($w$) and the relaxation channel ($W$).  
Despite the variation in $n$ and $\mu$,
the measured inelastic times are in close agreement; 
further evidence that $\tau_i$ depends only the lattice temperature (see Eq.~\ref{tau}).      
We argue in the next paragraph that the slightly higher values of $\tau_i$ in Devices 3 and 4 are 
caused by the narrower widths of the voltage probes. 

To use HETs in different studies, for example to map out the temperature profile $T_e(x)$ along the 
relaxation channel at lower temperatures, a number of design issues should be considered:
\newline (i) {\bf the width $W$ of the relaxation channel}  \  In non-local measurements,
a small fraction of  $I_H$  extends into the R-channel.
This is characterized by the non-local resistance $R_{nl}= V_t/ I_H$, 
where $V_t$ is the voltage measured across the R-channel at position $x$,  
which theory predicts\cite{Abanin2009} will decay exponentially 
down the R-channel as:
\begin{equation}
R_{nl} = F  R_{sq} \,  \exp(-x \, \pi/W),
\label{nl}
\end{equation}
where $F$ is a geometric factor of order unity.
For $W=60~\mu$m the voltage relaxation length is \mbox{$W/\pi \approx 19~\mu$m;}  
this has been confirmed experimentally by $V_t$ measurements at $x_1$,  $x_2$ and $x_3$ 
where $F \approx  0.5$, in agreement with a simple finite element model of the electrostatics.  
Therefore, the parasitic electron heating in the R-channel 
will extend over a length $W/(2 \pi)  \approx 10~\mu$m, 
which is smaller than any measured cooling length.  
The results from Device~3 show that $\cool$ is not changed when the width $W$ 
is reduced to $30~\mu$m, so we believe that parasitic heating is not affecting our results.    
\newline (ii) {\bf the width $w$ of the HET probes} \  The HET voltage probes 
are paths for heat to leak out of the relaxation channel.  
Finite element model simulations show that the measured $\cool$ and $\tau_i$ will be lowered as the width  $w$  increases, 
numbers typically being $5\%$ for $\cool$ and $10\%$ for $\tau_i$.  
This explains the trend that the measured $\tau_i$ is higher for Devices 3 and 4 compared to Devices 1 and 2. 
To make more accurate inelastic scattering time measurements, the smallest possible $w$ 
should be used, and the HETs should be run at more negative gate voltages. 
\newline (iii)  {\bf the spacing of the HET probes}  \ If the temperature is lowered to dilution fridge temperatures the 
exponential temperature profile, $ \Delta T(x) \sim \exp (-x/\cool)$, will be replaced by 
one that is determined by cooling by the Ohmic contacts on the HET voltage probes, 
as well as the Ohmic contact at the far end of the R-channel. 
To accurately measure the temperature profile in this limit requires HETs spaced evenly along the R-channel. 

In conclusion, we show how bar gate thermometers can be used to measure 
the cooling length $\cool$, wth a method that is applicable to other 2D materials which are gatable. 
Spatially separating the heating and cooling in non-local measurements is key to 
measuring the inelastic scattering time $\tau_i$.   
For $T_L=$2-5~K the measured $\tau_i$ has no dependence on carrier density $n$, 
but follows the temperature dependence for inelastic scattering by acoustic phonons. 
As the lattice temperature $T_L$ is reduced below 2~K cooling by the lattice diminishes, 
and cooling by the Ohmic contacts become more dominant. 
This crossover in cooling mechanism leads to a non-exponential temperature profile in the relaxation channel. 
Consistent with this crossover, for $T_L<2$~K there is no phonon drag contribution ($\Delta S^g$) to the 
measured thermopower of the bar gates, and the 2DEG resistivity $\rho(T)$
saturates to a constant value that is determined by the electron-impurity scattering.      

Measurements of $\Delta T(x)$ also provide information about the temperature gradient, $\nabla T(x)$, 
a quantity that determines the magnitude of the magneto-thermopower in a transverse magnetic field, 
which will be presented elsewhere.\cite{results}   
\vskip 2mm
\noindent {\bf Supplementary Material}  See supplementary information (SI) file for further information. 
\vskip 2mm
\noindent{\bf Acknowledgments}  This work was supported by EPSRC (UK) Programme Grant  EP/R029075/1. 
\vskip 2mm
\noindent {\bf Data availability:}  The data that support the findings of this study 
are available from the corresponding author upon reasonable request.


\begin{thebibliography}{10}

\bibitem{Lin2002}
J.~J. Lin and J.~P. Bird,
\newblock J. Phys.: Cond. Matt. {\bf 14}, R501 (2002).

\bibitem{appl98a}
N.~J. Appleyard, J.~T. Nicholls, M.~Y. Simmons, W.~R. Tribe, and M.~Pepper,
\newblock Phys. Rev. Lett. {\bf 81}, 3491 (1998).

\bibitem{Levitin2022}
L.~V. Levitin et~al.,
\newblock Nature Commun. {\bf 13}, 667 (2022).

\bibitem{Syme1989}
R.~T. Syme, M.~J. Kelly, and M.~Pepper,
\newblock J. Phys.: Cond. Matt. {\bf 1}, 3375 (1989).

\bibitem{Kurdak1995}
C.~Kurdak, D.~C. Tsui, S.~Parihar, S.~A. Lyon, and M.~Shayegan,
\newblock Appl. Phys. Lett. {\bf 67}, 386 (1995).

\bibitem{Chickering2009}
W.~E. Chickering, J.~P. Eisenstein, and J.~L. Reno,
\newblock Phys. Rev. Lett. {\bf 103}, 046807 (2009).

\bibitem{Mavalankar2013}
A.~Mavalankar et~al.,
\newblock Appl. Phys. Lett. {\bf 103}, 133116 (2013).

\bibitem{MC1969}
M.~Cutler and N.~F. Mott,
\newblock Phys. Rev. {\bf 181}, 1336 (1969).

\bibitem{Bakker2012}
F.~L. Bakker, J.~Flipse, and B.~J. van Wees,
\newblock J. Appl. Phys. {\bf 111}, 084306 (2012).

\bibitem{Sierra2015}
J.~F. Sierra, I.~Neumann, M.~V. Costache, and S.~O. Valenzuela,
\newblock Nano Lett. {\bf 15}, 4000  (2015).

\bibitem{Narayan2016}
V.~Narayan, M.~Pepper, and D.~A. Ritchie,
\newblock C. R. Phys. {\bf 17}, 1123 (2016).

\bibitem{Waissman2021}
J.~Waissman et~al.,
\newblock Nat. Nanotechnol. {\bf 17}, 166 (2021).

\bibitem{Billiald2015}
J.~Billiald et~al.,
\newblock Appl. Phys. Lett. {\bf 107}, 022104 (2015).

\bibitem{Rojek2014}
S.~Rojek and J.~K\"onig,
\newblock Phys. Rev. B {\bf 90}, 115403 (2014).

\bibitem{Massicotte2021}
M.~Massicotte, G.~Soavi, A.~Principi, and K.-J. Tielrooij,
\newblock Nanoscale {\bf 13}, 8376 (2021).

\bibitem{Henny1997}
M.~Henny et~al.,
\newblock Appl. Phys. Lett. {\bf 71}, 773 (1997).

\bibitem{Huard2007}
B.~Huard, H.~Pothier, D.~Esteve, and K.~E. Nagaev,
\newblock Phys. Rev. B {\bf 76}, 165426 (2007).

\bibitem{Arora1985}
V.~K. Arora and A.~Naeem,
\newblock Phys. Rev. B {\bf 31}, 3887 (1985).

\bibitem{Abanin2009}
D.~A. Abanin, A.~V. Shytov, L.~S. Levitov, and B.~I. Halperin,
\newblock Phys. Rev. B {\bf 79}, 035304 (2009).

\bibitem{results}
A.~K. Jain et~al.,
\newblock in preparation,
\newblock 2025.

\end{thebibliography}

\end{document}